# Electrochemical control of ferroelectricity in hafnia-based ferroelectric devices using reversible oxygen migration

Minghao Shao[1,†], Houfang Liu[1,†,*], Ri He[2,†], Xiaomei Li[3,†], Liang Wu[4,5], Ji Ma[5], Xiangchen Hu[6], Ruiting Zhao[1], Zhicheng Zhong[2], Yi Yu[6], Caihua Wan[3], Yi Yang[1], Ce-Wen Nan[7], Xuedong Bai[3,*], Tian-Ling Ren[1,*], X. Renshaw Wang[4,8,*]

[1]Institute of Microelectronics and Beijing National Research Center for Information Science and Technology (BNRist), Tsinghua University, Beijing 100084 China

[2]Key Laboratory of Magnetic Materials and Devices, Ningbo Institute of Materials Technology and Engineering, Chinese Academy of Sciences, Ningbo, 315201 China

[3]Beijing National Laboratory for Condensed Matter Physics, Institute of Physics, Chinese Academy of Sciences, Beijing 100190 China

[4]Division of Physics and Applied Physics, School of Physical and Mathematical Sciences, Nanyang Technological University, 21 Nanyang Link, 637371 Singapore

[5]Faculty of Materials Science and Engineering, Kunming University of Science and Technology, Kunming 650093 China

[6]School of Physical Science and Technology, ShanghaiTech University, Shanghai 201210 China

[7]School of Materials Science and Engineering, State Key Laboratory of New Ceramics and Fine Processing, Tsinghua University, Beijing 100084 China

[8]School of Electrical and Electronic Engineering, Nanyang Technological University, 50 Nanyang Ave, 639798 Singapore

†These authors contributed equally to this work.

*Corresponding author

Email: rentl@tsinghua.edu.cn (T.L.R.); houfangliu@tsinghua.edu.cn (H.F.L.); xdbai@iphy.ac.cn (X.D.B.); renshaw@ntu.edu.sg (X.R.W.)

## Abstract

Ferroelectricity, especially in hafnia-based thin films at nanosizes, has been rejuvenated in the fields of low-power, nonvolatile and Si-compatible modern memory and logic applications. Despite tremendous efforts to explore the formation of the metastable ferroelectric phase and the polarization degradation during field cycling, the ability of oxygen vacancy to exactly engineer and switch polarization remains to be elucidated. Here we report reversibly electrochemical control of ferroelectricity in $Hf_{0.5}Zr_{0.5}O_2$ (HZO) heterostructures with a mixed ionic-electronic $LaSrMnO_3$ electrode, achieving a hard breakdown field more than 18 MV/cm, over fourfold as high as that of typical HZO. The electrical extraction and insertion of oxygen into HZO is macroscopically characterized and atomically imaged *in situ*. Utilizing this reversible process, we achieved multiple polarization states and even repeatedly repaired the damaged ferroelectricity by reversed negative electric fields. Our study demonstrates the robust and switchable ferroelectricity in hafnia oxide distinctly associated with oxygen vacancy and opens up opportunities to recover, manipulate, and utilize rich ferroelectric functionalities for advanced ferroelectric functionality to empower the existing Si-based electronics such as multi-bit storage.

**Main Text:**
**Introduction**

Ferroelectricity, the property of exhibiting a reversibly-switchable electric polarization, is a rare but technologically useful functionality[1] that has been applied in a wide range of applications, such as non-volatile memories, sensors, optics, and electromechanical control[2,3]. Recently, the nanoscale ferroelectricity in doped $HfO_2$ is attracting attention owing to its unexpectedly strong ferroelectricity and ability to integrate into silicon electronics[4–8]. Even with the thickness down to two fluorite-structure unit cell dimension, a Zr-doped $HfO_2$ still possesses robust switchable electric dipoles[5]. The origin of the stable ferroelectricity is dominantly attributed to the formation of a metastable ferroelectric phase with the assistance of confining effect of surface/interface, dopants, residual strain and oxygen vacancies. Importantly, among them, understanding the role of oxygen concentration in the modulation of ferroelectric polarization and degradation during field cycling is crucial. Moreover, because of the complicated interplay between ionic charge and lattice degrees of freedom[9], it has always been extremely challenging to reversibly create multiple polarization states, heal fatigue or aged polarization state, especially in ultrathin ferroelectric $HfO_2$[10–12].

To achieve control of ferroelectric polarization states, the displacement of a suitable quantity of crucial cations and anions has to be controlled in a uniform, reversible, and accurate manner. In hafnia-based ferroelectric materials, re-activation of the collective ferroelectric order requires the inserted oxygen to delicately break the spatial inversion symmetry of the lattice structure and effectively form electric dipoles. Heterostructures utilizing a mixed ionic-electronic conductor[13,14] could be an attractive way to explore the control and healing of ferroelectricity. In a mixed conductor, not only does the electronic conduction meet the requirement for a metallic electrode but also the ionic conduction enables electrochemical coupling at the interface, with the insertion or extraction of oxygen ions. Moreover, because oxygen migration is able to induce structural phase transitions in epitaxially grown Zr-doped $HfO_2$ via a $LaSrMnO_3$ electrode[15], the tantalizing ion-lattice interactions in Zr-doped $HfO_2$ may allow the control as well as healing of the ferroelectricity in Zr-doped $HfO_2$.

Here, we report reversibly electrochemical control of ferroelectric polarization states in a heterostructure of $Hf_{0.5}Zr_{0.5}O_2$ (HZO) enabled by an electrode of the mixed conductor, $La_{0.67}Sr_{0.33}MnO_3$ (LSMO). By replacing a TiN electrode with a single crystalline mixed conductor LSMO, the remnant polarization ($P_r$) of HZO is enhanced by at least a factor of two and it remains stable even at a high electric field of 18 MV/cm, approximately close to intrinsic breakdown strength. Furthermore, multi-level control of the ferroelectricity is demonstrated induced by reversible and uniform insertion and extraction of oxygen ions under a certain operating voltage. The reversible migration of oxygen ions was observed *in situ* by scanning transmission electron microscopy.

**Results and discussion**

Fifty per cent Zr doped $HfO_2$ (HZO) films with a 15 nm nominal thickness were grown on $SrTiO_3$ (001) (STO) substrates with LSMO and TiN as bottom and top electrodes, respectively (see schematic in Fig. 1a). Polycrystalline HZO and single-crystalline LSMO were fabricated using atomic layer deposition (ALD) and pulsed laser deposition (PLD), respectively. The LSMO/HZO/TiN stack was subsequently subjected to rapid thermal annealing (see details in Methods). In our study, we chose the mixed ionic-electronic conductor, LSMO, as the bottom electrode. In addition, control samples of TiN/HZO/TiN were fabricated following the same protocol.

We characterized the structural properties of the LSMO/HZO/TiN heterostructure using high-resolution transmission electron microscopy (HRTEM). ALD-grown Zr-doped $HfO_2$ is a poly-crystalline ferroelectric material with multiple phases[16] and the *o*-phase[17] is typically responsible for ferroelectricity. Figure 1b confirms that HZO is polycrystalline and LSMO is monocrystalline (Figs. S1 and S2). Figs. 1c-f further show the coexistence of the ferroelectric *o*-phase and the paraelectric monoclinic *m*-phase. The structural results are, therefore, consistent with the existing literature.[7,17,18] Note that, although LSMO fabricated by PLD is single crystalline, the LSMO material can also be made polycrystalline or even amorphous using other growth techniques, such as sputtering, without jeopardizing its ionic conduction and integration into the modern microelectronic industry.

Bistable switching and polarization *vs*. electric field (*P-E*) hysteresis loops were obtained from positive-up-negative-down (PUND) measurements to rule out interference from the non-ferroelectric switching contribution (see experimental details in Fig. S3). Figure 2a shows the polarization of both LSMO/HZO/TiN and TiN/HZO/TiN heterostructures with fully saturated hysteresis. By replacing TiN with LSMO, the $P_r$ of LSMO/HZO/TiN is enhanced to ~24 $\mu C/cm^2$, much higher than the $P_r$ of ~14 $\mu C/cm^2$ in the TiN/HZO/TiN control sample and the reported $P_r$, ~10 $\mu C/cm^2$ of PLD-grown epitaxial HZO(111) films on the LSMO/STO(001) with similar thicknesses[19,20]. Moreover, the retention performance of the LSMO/HZO/TiN heterostructure reached up to $10^5$ s with only an approximately 5% loss (Fig. S4). By extrapolation, the loss after ten years was estimated to be 7% with a $P_r$ of 19.7 $\mu C/cm^2$.

Figure 2b shows the *P-E* loops of the LSMO/HZO/TiN and TiN/HZO/TiN heterostructures with electric fields higher than the typical breakdown electric field, $E_{BD}$, at 1 kHz (Figs. S5 and S6). The $E_{BD}$ can be characterized as the value of the electric field with a non-switching leakage current density over 10 A/cm$^{2}$[21]. Figure 2c shows that $E_{BD}$ of ferroelectric TiN/HZO/TiN is approximately 4 MV/cm. This value is quantitatively consistent with the reported value for Zr-doped $HfO_2$, which are approximitely 4 to 5 MV/cm[22], suggesting an inherently small margin during stable operation for saturated polarization switching with an applied electric field smaller than $E_{BD}$. A 4 MV/cm electric field can break down the TiN/HZO/TiN, but, remarkably, the ferroelectric properties of LSMO/HZO/TiN remain relatively stable after applying a 10 MV/cm electric field. Figure 2d compares the $E_{BD}$ of the LSMO/HZO/TiN heterostructure, the reported HZO, $SrBa_2Ta_2O_9$ (SBT) and $Pb(Zr,Ti)O_3$ (PZT), as a function of $P_r$. Without sacrificing much of the $P_r$ of ~20 µC/cm$^2$, the 10 MV/cm $E_{BD}$ of the LSMO/HZO/TiN heterostructure is over twice as high as that of typical HZO, and five times that of SBT and PZT ($E_{BD}$ < 2 MV/cm). Regarding fatigue and aging, Fig. 2d shows $E_{BD}$ values of 12 and 18 MV/cm for the LSMO/HZO/TiN heterostructure with retained $P_r$ of 14 and 9 µC/cm$^2$ (Fig. S7), respectively, also demonstrating that the LSMO electrode dramatically enhances the strength and robustness of the ferroelectricity in HZO.

The enhancement of $P_r$, good retention, and high $E_{BD}$ of ferroelectric LSMO/HZO/TiN are unexpected because a high field inevitably introduces fatigue and aging[23,24]. To gain insight into the exceptional ferroelectricity in LSMO/HZO/TiN, we conducted current density-electric field (*J-E*) measurements at 1 kHz with a high electric field. Figure 3a shows the *J-E* curve with driven electric fields of different ranges. Experimental details are shown in Fig. S8, and more electric field ranges are shown in Figs. S9 and S10. At a small driving field (*E* < 3 MV/cm), we observe clear polarization switching current peaks at approximately ±1.9 MV/cm (light-blue region in Fig. 3a) in all *J-E* curves, in agreement with the coercive field of the LSMO/HZO/TiN heterostructure (see Figs. S5 and S11). Interestingly, at an intermediate driving field (3 MV/cm < *E* < 5 MV/cm), an anomalous shoulder at ~3.8 MV/cm (light-orange region in Fig. 3a) shows an intensity approximately equal to the polarization switching current. Most importantly, the peak at ~3.8 MV/cm is absent in the TiN/HZO/TiN control samples (see Fig. S12), proving that an interfacial interaction occurs between the LSMO and HZO. Moreover, at high driving fields (*E* >5 MV/cm), the background leakage current of the LSMO/HZO/TiN heterostructure rises sharply due to the formation of conductive filaments likely produced by soft breakdown at the HZO grain boundaries[25]. Note that, although the high-field breakdown effect in LSMO/HZO/TiN is qualitatively similar to that in TiN/HZO/TiN, LSMO/HZO/TiN exhibits two distinctly different features: (i) a higher breakdown field, and (ii) robust ferroelectricity after applying an electric field above 4 MV/cm, where TiN/HZO/TiN loses its ferroelectricity.

We suppose that the technical advantages of LSMO/HZO/TiN and the *J-E* peak in the intermediate electric field range originate from controllable and reversible mixed-conductor-assisted atomic oxygen migration at the LSMO/HZO interface. Negatively-charged oxygen ions, which are equivalent to positively-charged oxygen vacancies, can be displaced by an electric field, leading to the migration of oxygen ions. We confirm this by using a comprehensive model combining first-principles calculations and the dynamic simulation of oxygen ion movement[26]. Figure 3b shows our theoretical fitting of the *J-E* characteristics based on three distinct processes from low to high electric fields: (i) ferroelectric switching (green line in Fig. 3b), (ii) oxygen ion migration (orange line in Fig. 3b), and (iii) conductive filament formation (blue line in Fig. 3b).

Especially at intermediate electric fields (3 MV/cm < *E* < 5 MV/cm), the predominant effect is that the negatively-charged oxygen ions reversibly migrate and carry electron current. We employed an electrothermal model to simulate the diffusive migration of oxygen ions under voltage pulses and to fit the *J-E* characteristics of the LSMO/HZO/TiN heterostructure. The oxygen ion diffusion continuity equation, current continuity equation, and Joule heating equation were self-consistently solved. In the simulation, we used an LSMO/HZO configuration and the thicknesses of HZO and LSMO used in the experiment. Figure S13 shows the evolution of the oxygen ion and vacancy distribution during the application of triangular voltage pulses from 0 to 5 MV/cm. Under a positive electric field, the negatively-charged oxygen ions migrate towards the positive potential (HZO region) and away from the LSMO electrode (see simulated results in Fig. S13a-c and schematic in Fig. 3c), leading to a filling of the oxygen ions in HZO. When the electric field is large enough to overcome the activation energy ($E_a$) of oxygen vacancy diffusion, which is the energy barrier preventing the oxygen ions from migrating and is calculated by first-principles calculations (see Supplementary Notes), the particle flux of oxygen ions across the LSMO/HZO interface increases first with increasing electric field intensity or duration. With the exhaustion of oxygen vacancies in HZO, the particle flux decreases with increasing intensity or duration of the electric field. Consequently, with a further increase in the electric field, the migration of oxygen ions stops, and the current response becomes time-invariant. This process forms a peak at a certain electric field, called the redox peak. When opposite triangular voltage pulses were applied from 0 to -5 MV/cm, the oxygen ions migrated back into the LSMO under a negative electric field (see simulated results in Fig. S13d and schematic in Fig. 3d). This is a process equivalent to the electrical creation of oxygen vacancies in HZO, because the oxygen ions have opposite charges to the oxygen vacancies. The simulated current value of the redox peak occurs at 3.4 MV/cm, which is quite consistent with the experimentally observed peak in Fig. 3a. Moreover, because the flux of oxygen ions is also dependent on the duration of the electric field, the oxygen migration model was further confirmed by the experimental *J-E* characteristics as a function of the frequency of scanning the driving electric field. We found that the electrochemical peak shifts to lower fields with a lower sweep rate, because a longer duration exhausts the available oxygen ions (see Figs. S14 and S15).

More excitingly, the controllable insertion and extraction of oxygen ions from HZO using a mixed ionic-electronic conductor provide an opportunity to reversibly control the magnitude of the polarization of HZO to multiple levels.

Initially, upon conventional application of an electric field lower than the $E_{BD}$ of the heterostructure and the $E_a$ of oxygen vacancy diffusion, $P_r$ of LSMO/HZO/TiN can be switched to ~+22 or ~-22 µC/cm$^2$, which are defined as the "+1" and "-1" states, respectively (see Fig. 3e). However, the bottleneck to achieving control in ferroelectricity is the re-activation of the ferroelectric polarization state via oxygen migration (see the breakdown of ferroelectricity in TiN/HZO/TiN in Fig. S16). Using the mixed ionic-electronic conductor, we performed both the electric pulse intensity- and duration-dependent modulation of oxygen migration to modulate the polarization to different levels. First, Fig. 3f shows that, by applying square pulses of +3.5 MV/cm for 50, 100, and 1000 ms and +4.5 MV/cm for 1000 ms, the $P_r$ of HZO decreased from 22 to 18, 16, 14, and 12 µC/cm$^2$, respectively. If the initial polarization state is defined as "+1" or "-1" and the new polarization states are "+2", "+3", "+4", and "+5", then Fig. 3f shows that multiple polarization states are established by the external electric field. For each ±n polarization state, there is a unique, reversible hysteresis loop. Examples are shown in Fig. 3e and Fig. S17a. Moreover, multi-level polarization states can also be achieved by tuning the duration of the pulse. Figure 3g shows that, with a fixed magnitude of the pulse of +3.5 MV/cm, the $P_r$ of HZO also decreases from ~22 to ~18 and ~16 µC/cm$^2$ by applying 50 and 100 ms, respectively. Following the same protocol, a controllable state from the "+1" to "+2" and "+3" states is achieved reversibly. The full hysteresis loops of the polarization states are shown in Fig. S17b. Hence, we have experimentally demonstrated controllable ferroelectricity enabled by the insertion and extraction of oxygen ions from the mixed conductor in a quantitatively controllable, electrically reversible, and spatially uniform manner.

At a fundamental level, because multi-levels of the polarization state were achieved by the extraction and insertion of oxygen vacancies, our results also demonstrate a direct correlation between the oxygen vacancy and the ferroelectric polarization. To provide insights into the correlation between oxygen and ferroelectricity in HZO, we performed *in situ* high-angle annular dark field (HAADF) electron energy-loss spectroscopy (EELS) under an electric field[27]. Figure 4a shows the schematic and results of the *in situ* scanning transmission electron microscopy (STEM) real-time measurements under triangular voltage pulses with a magnitude of 10 V and a fixed duration applied through the TiN top and LSMO bottom electrodes. We chose eleven positions, labelled from 1 to 11, on the LSMO side of the heterostructure to measure the valence changes of Mn and O (see Fig. 4b). The HAADF-EELS spectra were measured instantaneously after the application of short triangular voltage pulses between the metal tip and conducting LSMO electrode. In Fig. 4b, the EELS data confirm the migration of oxygen ions from the LSMO film to HZO when positive voltage pulses are applied, and the re-oxidation of LSMO after the application of negative voltages. The $L_3$ and $L_2$ peaks, due to the transitions from $2p_{3/2}$ and $2p_{1/2}$ core states to $3d$ unoccupied states, respectively, are localized on the excited manganese ions. The position of the $L_3$ peak tends to shift to a higher energy as the valence of manganese ions increases, and *vice versa*[28,29]. As shown by the red arrow in Fig. 4b, the $L_3$ peak shifts along the lower energy position after a positive voltage, suggesting that the valences of the manganese ions decrease by the oxygen ions dissociating from LSMO and migrating into HZO. Additionally, the initial state can be restored when a negative *in situ* electric field is applied. Considering the energy difference between the Mn $L_3$ edge and O $K$ edge in Fig. 4c and the Ti $L$ edge in Fig. S18, the energy difference, $\Delta E$, between Mn $L_3$-O $K$ ($\Delta$ Mn-O) shows a monotonic increase with the Mn nominal oxidation state[30,31]. Figure 4d shows the $\Delta E$ (Mn $L_3$ - O $K$) from the LSMO to the HZO (marked from 1 to 11 in Fig. 4a) in different states. $\Delta E$ (Mn $L_3$-O $K$) decreases when applying a positive voltage, indicating a decrease in the valence of Mn. Notably, Fig. 4b, d also shows that, after a negative voltage, the Mn valence almost returns to its initial state, microscopically confirming the reversible and non-volatile migration of oxygen ions enabled by the LSMO.

In addition, the correlation between oxygen vacancies and ferroelectricity can also be explained by first-principles calculations. The oxygen vacancy stabilizes the *o*-phase of HZO, and the concentration of the oxygen vacancy determines the magnitude of spontaneous polarization on HZO (Fig. S19). Hence, considering all the results, microscopic oxygen ion migration, macroscopic controllable ferroelectric polarization states, and first-principles calculations, our work demonstrates that the controllable ferroelectricity is achieved by the electric-field-controllable migration of atomic oxygen. Furthermore, because most attention has been focused on the impact of strain and crystalline phases on the ferroelectricity of HZO, the observed direct correlation between oxygen and ferroelectricity provides previously missing insight into the origin of the ferroelectricity of HZO.

## Conclusion

In conclusion, we have demonstrated reversible control of polarization state in HZO from both macroscopic and microscopic perspectives by taking advantage of the mixed ionic-electronic conductor LSMO. Initial, enhanced, reduced, and re-activated ferroelectricity with reversible control was achieved in a reversible and controllable manner. Microscopic imaging of the reversible oxygen vacancy diffusion across the HZO/LSMO interface was captured in real-time by *in situ* STEM-EELS and demonstrates the correlation between oxygen vacancies and ferroelectricity in HZO. Our study provides insight into the origin of ferroelectricity in HZO, but the control approach could be easily extended to other ferroelectric operations (*e.g.,* creation, manipulation, and utilization) and to different oxide functionalities (*e.g.,* ferromagnetism and superconductivity), contributing to future electronic applications and multifunctional electronics.

## Methods

### Sample growth

$La_{0.67}Sr_{0.33}MnO_3$ (LSMO) (20 nm) was grown on a (001)-$SrTiO_3$ (STO) substrate using pulsed laser deposition (PLD) at an oxygen partial pressure ($P_{O2}$) of 20 Pa at 700 °C. After deposition, the films were post annealed at a deposition temperature and $P_{O2}$ of 20000 Pa for 1 hour. Then, the sample was cooled down at the same pressure (10 mTorr) at a speed of 10 °C/min. All the target materials were vaporized by irradiation with an excimer laser (λ = 248 nm) operating at a repetition rate of 2 Hz and laser fluence of 1.8 J/cm$^2$.

HZO films (15 nm) were grown by ALD at 200 °C with a $HfO_2$ and $ZrO_2$ cycle ratio of 1:1, using Tetrakis (ethylmethylamido) hafnium (IV) [TEMAH; $Hf(NCH_3C_2H_5)_4$], Tetrakis (ethylmethylamido) zirconium (IV) [TEMAZ; $Zr(NCH_3CH_5)_4$], and deionized water ($H_2O$) as the Hf-precursor, Zr-precursor, and oxygen source, respectively. Round TiN top electrodes with a diameter of 100 μm and a thickness of 20 nm were pre-patterned by a hard mask and sputtered in an argon environment at 2 mTorr.

After deposition, all the devices were annealed in ambient $N_2$ at 600 °C for 30 s by using rapid thermal annealing (RTA) for crystallization.

**TEM characterization**

The cross-sectional transmission electron microscopy (TEM) specimen was prepared by using a focus ion beam (FIB) in JEOL JIB-4700F. Ion-beam-induced carbon deposition was used to protect the top surface of the film. Thick lamella were transferred to a TEM grid using an *ex situ* lift out system. In the thinning process, the focused ion beam was operated at 30 kV and decreased to 5 kV for final thinning and the removal of surface amorphous layers. The final thickness of the lamella was less than 100 nm. TEM imaging, high-resolution TEM (HRTEM), and energy-dispersive X-ray spectroscopy (EDS) were performed on a JEOL JEM-F200 microscope, operated at 200 kV with a field-emission gun and a Gatan Rio16 camera. To increase the signal-to-noise ratio of the HRTEM images, the average background subtraction filter (ABSF) was applied to the HRTEM images in DigitalMicrograph (http://www.dmscripting.com/hrtem_filter.html).

***In-situ* STEM**

*In-situ* STEM experiments were carried out using an aberration-corrected JEOL ARM 300F at 300 kV in STEM mode. The STEM-EELS spectra were recorded using a JEOL ARM 300F with a double-tilt holder provided by the ZEPTools Technology Company. Electrical contacts to the LSMO side of the sample were made by gluing the sample to the half molybdenum ring with silver glue. The amorphous carbon on the W tip, as a good conductor, reduces the strain caused by the tip while ensuring effective contact.

**Positive-up-negative-down (PUND) measurement**

A positive-up-negative-down (PUND) measurement was carried out by applying 5 triangular pulses, as shown in Fig. S3, to obtain the remnant hysteresis loops of ferroelectric HZO. The first negative pulse as a preset pulse was used to switch all ferroelectric domains into the negative direction. Then, the positive switched pulse was applied and the switching half loop was measured. Note that this half loop contained both the ferroelectric switching part and the non-switching part. After a delay period, the positive unswitched pulse was applied to measure the non-switching part and obtain the non-switching half loop. The next two negative pulses were used to complete the other half loops. Then the remnant hysteresis loop could be calculated using the switching loop minus the non-switching loop, which contained only the ferroelectric switching part. In this experiment, the width and magnitude of the pulses were set to 0.5 ms and 3 MV/cm and the delay period was 1 s.

## Ferroelectric characterization

The remnant polarization and switching current of the HZO films were measured by Multi-Ferroic-II ferroelectric analyzer. The samples were fully waken up by square pulses with $10^4$ cycles, at 3 MV/cm at 1 kHz before the measurement. Bipolar triangular waves at 1 kHz were used in all polarization – electric field (*P-E*) and current density – electric field (*J-E*) tests in this study except as stated separately. The PUND method was used in the remnant polarization measurement.

## *P-V* and *I-E* measurement

A standard bipolar triangular waveform amplitude of $V_{max}$ and duration in milliseconds ($t_w$) was applied to measure the *P-V* and *I-E* curves, as shown in fig. S8. The waveform begins at 0 V and steps to a maximum value of the assigned voltage. Then, it proceeds to step to the negative of the assigned maximum. Last, the waveform steps back to zero volts. A voltage waveform is applied to the sample in a series of voltage steps. At each step, the polarization can be converted from the integrated value of the captured current. The polarization can be obtained by $P = CV/S = \int C_{sense} dV / S$. The current loops can also be inferred directly from the sensing voltage.

## Data and materials availability

All data is available in the main text or the supplementary materials.

## Acknowledgments

X.R.W. is grateful to J. M. D. Coey for reviewing the manuscript. We thank K. Huang for the useful insights. Funding: T.L.R. acknowledges support from the National Key R&D Program (2016YFA0200400) and the National Natural Science Foundation (U20A20168, 51861145202) of China. C.W.N. acknowledges support from the Basic Science Center Program of NSFC (grant No. 51788104). X.R.W. acknowledges support from Academic Research Fund Tier 2 (Grant No. MOE-T2EP50210-006), the Singapore National Research Foundation (NRF) under the Competitive Research Programs (CRP Grant No. NRF-CRP21-2018-0003), and the Agency for Science, Technology and Research (A*STAR) under its AME IRG grant (Project No. A20E5c0094). X.D.B. acknowledges support from the NSF (Grant Nos. 51991344 and 21773303) and CAS (Grant Nos. XDB33030200 and ZDYZ2015-1) of China. Z.C.Z. acknowledges support from the National Key R&D Program of China (2017YFA0303602) and the Key Research Program of Frontier Sciences of CAS (Grant No. ZDBS-LY-SLH008).

## Author contributions

T.L.R., H.F.L., and X.R.W. designed and directed the project. M.H.S carried out the experiment with the help of H.F.L., R.T.Z., L.W., and J.M. X.M.L and X.D.B. performed in situ STEM measurements and analyzed the data. R.H. and Z.C.Z. performed the first-principles calculations. X.H. and Y.Y. prepared the FIB samples and performed HRTEM & STEM-EDS measurements. X.R.W. extensively helped in understanding the structure and symmetry of the films. H.F.L., Z.C.Z., C.H.W., X.D.B., T.L.R., and X.R.W. analyzed the data. H.F.L., R.H., X.M.L. and X.R.W. wrote the manuscript with feedback from all co-authors. All co-authors discussed the results and commented on the manuscript.

**Competing interests**

The authors declare no competing interests.

**Supplementary Information**

Supplementary Figures (*1-19*)

Table S1

References (*1-20*) and notes (*1-6*)

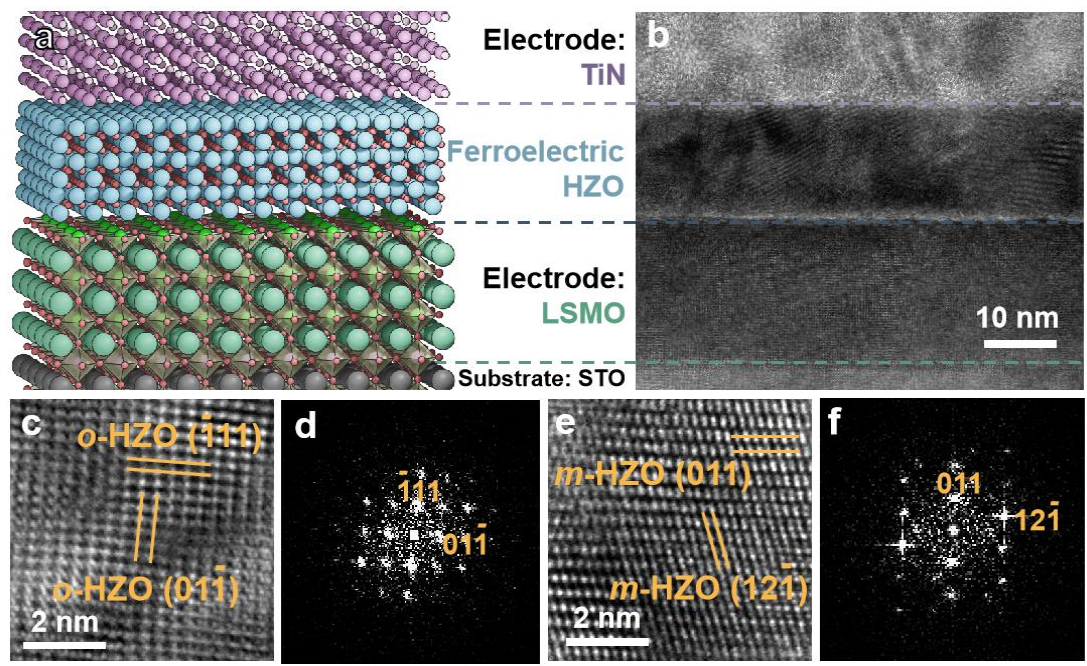

**Fig. 1 Structural characterization of the LSMO/HZO/TiN heterostructure.** (**a** and **b**) Schematic (**a**) and corresponding HRTEM image (**b**) of the TiN/HZO/LSMO heterostructure. LSMO and HZO are monocrystalline and polycrystalline, respectively. Scale bar is 10 nm. (**c** and **d**) HRTEM and corresponding Fourier transform (FT) patterns of the *o*-phase in polycrystalline HZO. Scale bar is 2 nm. (**e** and **f**) HRTEM and corresponding FT patterns of the *m*-phase in polycrystalline HZO. Scale bar is 2 nm.

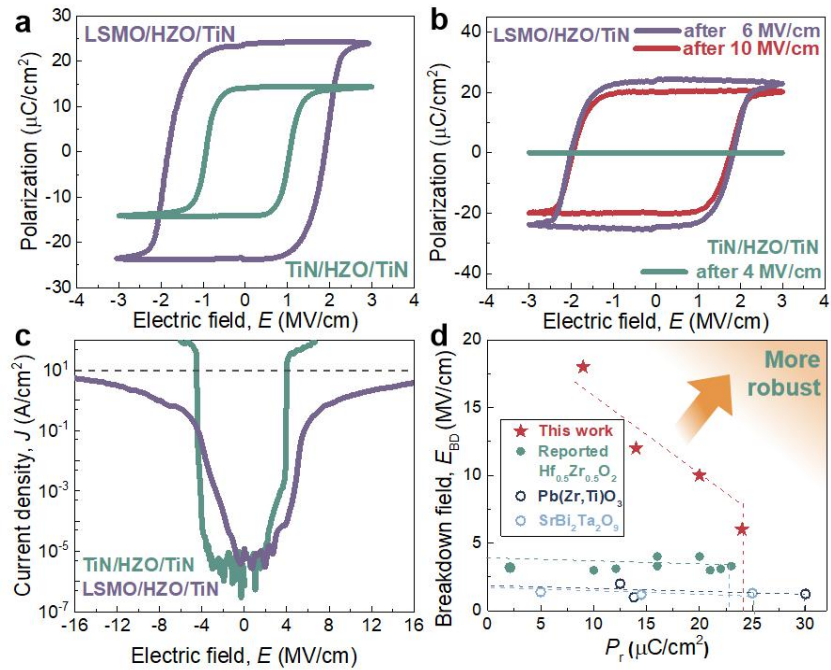

**Fig. 2 Enhancement, retention, and robustness of ferroelectricity in HZO heterostructures. a** The initial state of HZO polarization in LSMO/HZO/TiN and TiN/HZO/TiN. **b** Robust polarization of LSMO/HZO/TiN after applying high electric fields of 6 and 10 MV/cm (PUND measurement), contrasted with the vanishing ferroelectricity in TiN/HZO/TiN after applying a high electric field of 4 MV/cm. **c** Leakage current density as a function of the electric field for TiN/HZO/TiN and LSMO/HZO/TiN heterostructures. Note that the breakdown field ($E_{BD}$) can be determined by the electric field value at a current density of 10 A/cm². **d** Benchmark of $E_{BD}$ versus remanent polarization ($P_r$) for reported HZO, Pb(Zr,Ti)O$_3$ (PZT), SrBi$_2$Ta$_2$O$_9$ (SBT) films, and the LSMO/HZO/TiN heterostructure of the present work.

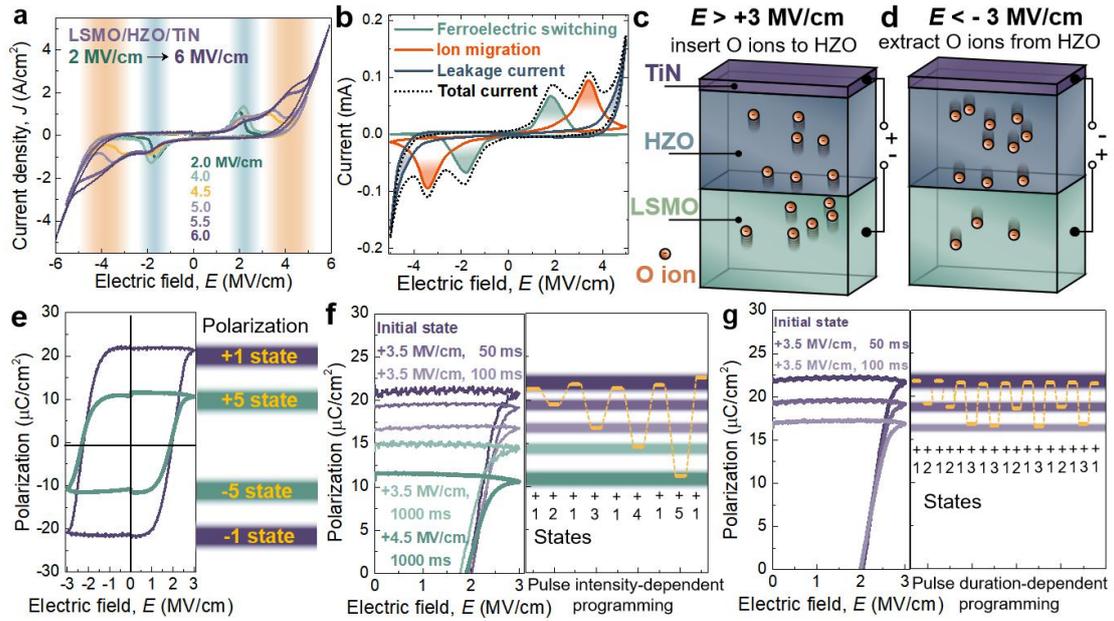

**Fig. 3 Reversibly control of polarization states in HZO. a** $J$-$E$ characteristics of LSMO/HZO/TiN under various electric fields. **b** Theoretical simulation of the dynamic processes of the controllable HZO, namely ferroelectric switching, ion migration, and leakage current. (**c** and **d**) Schematics of the reversibly-migratable oxygen ions in HZO in contact with a mixed conductor, LSMO. **e** Definition of representative "±1" and "±5" polarization states due to the switchable polarization of the HZO in the LSMO/HZO/TiN heterostructure. (**f** and **g**) Pulse intensity- and duration-dependent control of HZO polarization to reversible multiple polarization states with the help of the mixed conductor, LSMO.

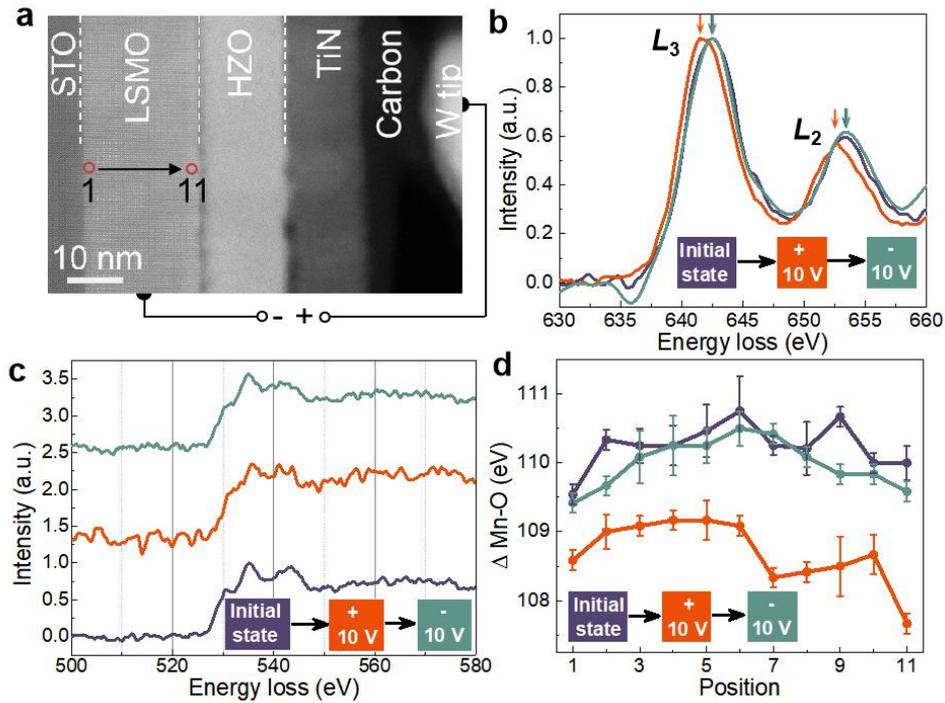

**Fig. 4 *In situ* STEM results of oxygen migration. a** STEM-HAADF image of the W tip/TiN/HZO/LSMO/STO structure. Scale bar is 10 nm. (**b** and **c**) STEM-EELS of the manganese $L_{2,3}$ edges and oxygen $K$ edges for the initial state, after positive voltage, and after negative voltage. **d** The energy difference between Mn $L_3$ and O $K$ edges - ΔE (Mn $L_3$-O $K$) - in different locations.

# Supplementary Information

# Electrochemical control of ferroelectricity in hafnia-based ferroelectric devices using reversible oxygen migration

Minghao Shao[1,†], Houfang Liu[1,†,*], Ri He[2,†], Xiaomei Li[3,†], Liang Wu[4,5], Ji Ma[5], Xiangchen Hu[6], Ruiting Zhao[1], Zhicheng Zhong[2], Yi Yu[6], Caihua Wan[3], Yi Yang[1], Ce-Wen Nan[7], Xuedong Bai[3,*], Tian-Ling Ren[1,*], X. Renshaw Wang[4,8,*]

[1]Institute of Microelectronics and Beijing National Research Center for Information Science and Technology (BNRist), Tsinghua University, Beijing 100084 China

[2]Key Laboratory of Magnetic Materials and Devices, Ningbo Institute of Materials Technology and Engineering, Chinese Academy of Sciences, Ningbo, 315201 China

[3]Beijing National Laboratory for Condensed Matter Physics, Institute of Physics, Chinese Academy of Sciences, Beijing 100190 China

[4]Division of Physics and Applied Physics, School of Physical and Mathematical Sciences, Nanyang Technological University, 21 Nanyang Link, 637371 Singapore

[5]Faculty of Materials Science and Engineering, Kunming University of Science and Technology, Kunming 650093 China

[6]School of Physical Science and Technology, ShanghaiTech University, Shanghai 201210 China

[7]School of Materials Science and Engineering, State Key Laboratory of New Ceramics and Fine Processing, Tsinghua University, Beijing 100084 China

[8]School of Electrical and Electronic Engineering, Nanyang Technological University, 50 Nanyang Ave, 639798 Singapore

**Supplementary Notes**

**First-principles calculation**

The activation energy of oxygen vacancy migration was determined by first-principles calculations based on density functional theory using a plane-wave basis set with a cutoff energy of 500 eV, as implemented in the Vienna *ab initio* simulation package (VASP) (*1–3*). Electron exchange and correlation effects were described within the generalized gradient approximation (GGA) (*4*). The cutoff energy for the plane-wave basis was chosen to be 500 eV. The polar orthorhombic (*o*-) phase with a Pca2$_1$ space group and the nonpolar monoclinic (*m*-) phase with a P2$_1$/c space group were considered in our calculation. The oxygen vacancy in *o*- and *m*-phase HfO$_2$ was modeled through periodically $2 \times 1 \times 2$ and $2 \times 2 \times 2$ supercells containing 32 Hf and 64 oxygen atoms. The Brillouin zone was sampled using a 4×4×4 Monkhorst-Pack *k*-point mesh for supercell relaxations. The Born effective charges along the *c*-axis were obtained by density functional perturbation theory: $Z_{Hf}^* = 4.86$, $Z_{O1}^* = -2.15$, $Z_{O2}^* = -2.61$ (O1 and O2 represent threefold and fourfold coordinated oxygens). These effective charge values agree well with previous calculations in the literature (*5*). The activation energies of oxygen vacancy, $E_a$, movement between the nearest-neighbor sites in La$_{0.67}$Sr$_{0.33}$MnO$_3$ (LSMO), LaMnO$_3$ and HfO$_2$ were calculated with a nudged elastic band (NEB) method (*6*).

**Dynamic oxygen vacancy movement simulations**

In the model, the oxygen vacancy is considered to be a positively charged (+2*e*) particle, which has the opposite charge of an oxygen ion. Initially, the oxygen vacancies had a uniform distribution in the HZO films (shown in Fig. S13a) (*7*). Hence, oxygen vacancies would move under an electric field. To describe the movement of oxygen vacancies upon the application of triangular electric field pulses, the time-dependent evolution of the oxygen vacancy concentration ($N_{Vo}$) can be described by the continuity equation as follows (*8*):

$$\frac{\partial N_{Vo}(r,t)}{\partial t} = \nabla \cdot (D\nabla N_{Vo} - vN_{Vo} + DS\nabla T N_{Vo}) \quad (1)$$

where *D* is the diffusion coefficient, *v* is the drift velocity, and *S* is the Soret coefficient for thermal diffusion. The expressions of the three coefficients are:

$$D = \frac{1}{2}a^2 f \exp\left(\frac{-E_a}{kT}\right) \quad (2)$$

$$v = af \exp\left(\frac{-E_a}{kT}\right) \sinh\left(\frac{qaE}{kT}\right) \quad (3)$$

$$S = \frac{-E_a}{kT^2} \quad (4)$$

where another parameter of *a* is the hopping distance in the lattices and *a* = 0.25 nm in HZO and 0.1 nm in LSMO, $E_a$ is the activation energy for oxygen vacancy movement in the lattices, $E_a$ is 0.7 eV for LSMO and 0.45 eV (positively charged oxygen vacancy) for HZO, and *f* is the escape-attempt frequency, and the value is 10$^{12}$ Hz in this model. The $E_a$ of LSMO is quantitatively similar to that of LaMnO$_3$ based on first-principles calculations. The drift-diffusion continuity equation can be solved by coupling with the current continuity equation:

$$\nabla \cdot J = \nabla \cdot \sigma E = -\nabla \cdot \sigma \nabla \psi = 0 \tag{5}$$

where $J$ is the current density, $\sigma$ is the conductivity, $E$ is the electric field and $\Psi$ is the electric potential. The room temperature conductivities of HZO and LSMO were set to 0.001 and 67 Scm$^{-1}$. In HZO, the conductivity was assumed to increase by $N_{VO}$.

The temperature was calculated using the heat equation:

$$-\nabla \cdot k \nabla T = J \cdot E \tag{6}$$

where $k$ is the thermal conductivity and was set to 1.1 and 2 W/mK for HZO and LSMO, respectively.

Equations (1), (5), and (6) were self-consistently solved to obtain the distribution of oxygen vacancy concentration and *I-E* characteristics as shown in Fig. 3b and Fig. S15.

The leakage currents of 0 to 5 MV/cm and 5 to 0 MV/cm were fitted by two equations:

$$I = 2.33 \times 10^{-5} \exp\left(\frac{-E}{-0.567}\right) + 3.53 \times 10^{-3} \tag{7}$$

$$I = 5.96 \times 10^{-9} \exp\left(\frac{-E}{-0.291}\right) + 3.69 \times 10^{-3} \tag{8}$$

**Phase-field simulation**

In the phase-field model, the evolution of polarization towards equilibrium is described by the time-dependent Ginzburg–Landau (TDGL) equation:

$$\frac{\partial P_i(r,t)}{\partial t} = -L \frac{\delta F}{\delta P_i(r,t)}, \quad i = 1, 2, 3 \tag{9}$$

where r is the spatial position and L is the kinetic relaxation coefficient. The total free energy F includes the three relevant energetic contribution items and can be written as follows (*9*):

$$F = \int_V (f_{bulk} + f_{grad.} + f_{elec.}) dV \tag{10}$$

where $f_{bulk}$, $f_{grad.}$, and $f_{elec.}$ are the bulk, gradient, and electrostatic contributions, respectively. The expressions of the three items are:

$$f_{bulk} = \alpha P^2 + \beta P^4 + \gamma P^6 \tag{11}$$

$$f_{grad.} = \frac{g(\nabla P)^2}{2} \tag{12}$$

$$f_{elec.} = -E^a P \tag{13}$$

where $\alpha$, $\beta$, and $\gamma$, are the expansion coefficients, $g$ is the gradient coefficient, and $E^a$ is applied electric field. The values of these expansion coefficients determine the thermodynamic behavior of the ferroelectric phase transition as well as the bulk ferroelectric properties.

The discretized two-dimensional simulation size was 256 $dx$ × 256 $dx$, with a grid mesh spacing of $dx$ = 0.1 nm. According to recent studies (*10–12*), the polarization direction in the polar *o*- phase is along the *c*-axis. Therefore, we assumed that the $P_x = 0$ and $P_y \neq 0$. The periodic boundary condition was imposed in two directions, and a short-circuit electric boundary condition was imposed in our simulation. The dynamics of the $P_y$ dependence on the applied triangular periodic electric field pulses (see Fig. S11) are obtained by numerically solving the TDGL equations using the semi-implicit Fourier spectral method. The expansion parameters of HZO used in the model were taken based on experimental data by the approach in Ref. (*13*) and all the parameters are listed in Table S1.

**Dynamic ferroelectric switching characteristics of HZO film**

To describe the detailed dynamical ferroelectric switching characteristics in ferroelectric HZO thin films under the application of triangular negative and positive low electric field pulses, we performed numerical simulations by using a two-dimensional phase-field model. In the phase-field model, the evolution of polarization switching can be obtained by solving the TDLG equation. Fig. S11 shows the simulated polarization - electric field loop and ferroelectric switching current - electric field characteristics of a single domain HZO thin film. $P_r$ is ~20 μC/cm$^2$, and the peak switching current is at 1.8 MV/cm, which is consistent with the experiment.

**Redox peak as a function of the scanning field frequency and strength**

We also found that with a faster sweep rate and a larger electric field, more peaks shifted to the right, which is in agreement with the experimental *I-E* curves. Experimentally, as shown in Fig. S14b, the anomalous peaks in the moderate electric field shift toward high electric fields as the scanning frequency increases. This change is consistent with the theoretical calculation of the dynamics of oxygen migration. The slower the scanning frequency of the electric field is, the longer the duration that the electric field stays at a certain electric field, and the easier oxygen atoms can migrate under the electric field. Additionally, a lower electric field is needed to form the redox peak.

**The effect of oxygen vacancies on ferroelectric properties**

In parallel to the theory of intrinsic ferroelectricity from the polar *o*-phase, several experimental and theoretical reports have indicated that oxygen vacancies also play a significant role in the evolution of ferroelectric properties or even induce ferroelectricity in HfO$_2$-based films (*14–16*). For example, it is found that the nonpolar *m*-phase and tetragonal phase transform to the polar *o*-phase when oxygen vacancies migrate into HfO$_2$ (*17, 18*), which is thought to be a potential origin of the wake-up effect. Glinchuk *et al.*'s theoretical research indicates that oxygen vacancies could induce the electric dipoles in $HfO_2$, and the homogeneous distribution of electric dipoles under electric cycling is responsible for reversible ferroelectric polarization (*16*). However, the underlying mechanism for the effect of oxygen vacancies on ferroelectric properties looks more unclear because it is very hard to visualize the oxygen vacancy distribution in polycrystalline $HfO_2$ during electric cycling. Our density functional theory (DFT) calculation indicates that the oxygen vacancy reduces the energy difference between the polar *o*-phase and the nonpolar *m*-phase (see Fig. S19a), which agrees with previous theoretical studies (*19, 20*), although the energy of the polar *o*-phase is still higher than that of the *m*-phase. This implies that the injection and storage of oxygen vacancies in HZO may promote the formation of the polar *o*-phase and result in the creation of ferroelectric properties. In addition, the introduction of oxygen vacancies can also induce a lattice distortion in the polar *o*-phase. Consequently, these vacancies can enhance the spontaneous polarization magnitude of the *o*-phase and lower the polarization switching barrier energy (Fig. S19b).

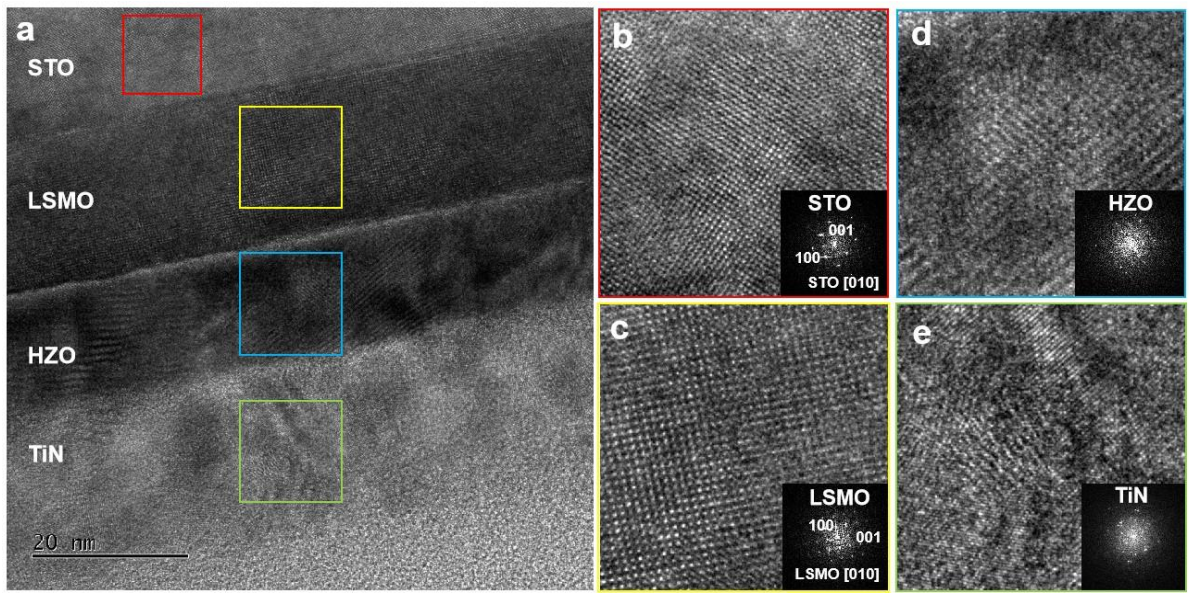

**Figure S1. Cross-sectional TEM images of the LSMO 20 nm/HZO 15 nm/TiN 20 nm heterostructure on the SrTiO₃ (STO) substrate. a** High-resolution transmission electron microscopy (HRTEM) image of the LSMO/HZO/TiN heterostructure. (**b-e**) Magnified views and corresponding Fourier transform (FT) patterns of the core layers of the heterostructure, namely (**b**) STO substrate, (**c**) LSMO, (**d**) HZO, and (**e**) TiN layer.

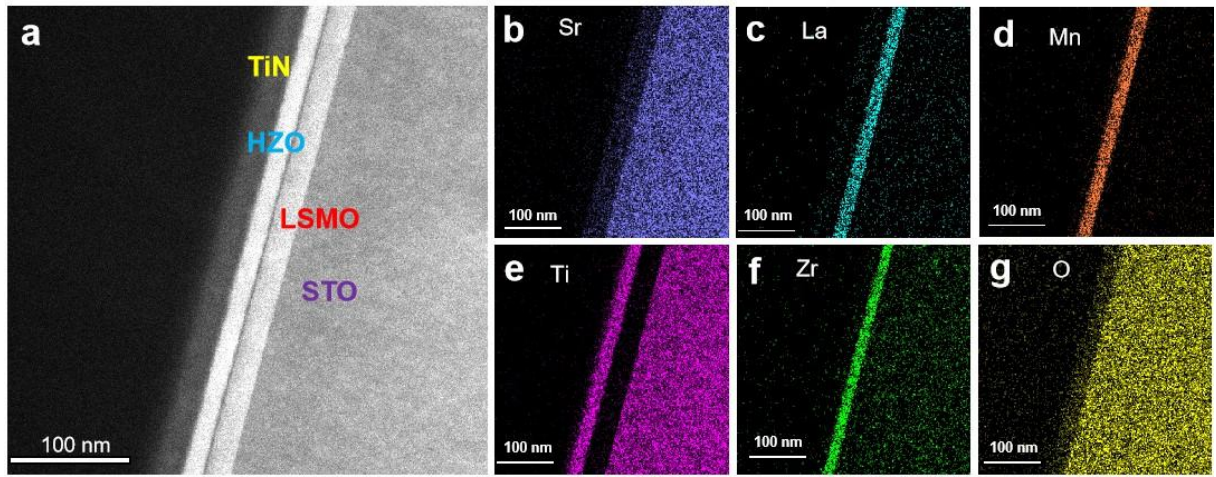

**Figure S2. Characterizations of the LSMO/HZO/TiN heterostructure. a** High-angle annular dark field (HAADF) image. (**b-g**) Energy-dispersive X-ray spectroscopy (EDS) maps of Sr, La, Mn, Ti, Zr, and O in the LSMO/HZO/TiN heterostructure. The results indicate that the LSMO/HZO/TiN heterostructure has structurally and chemically sharp interfaces and negligible interdiffusion.

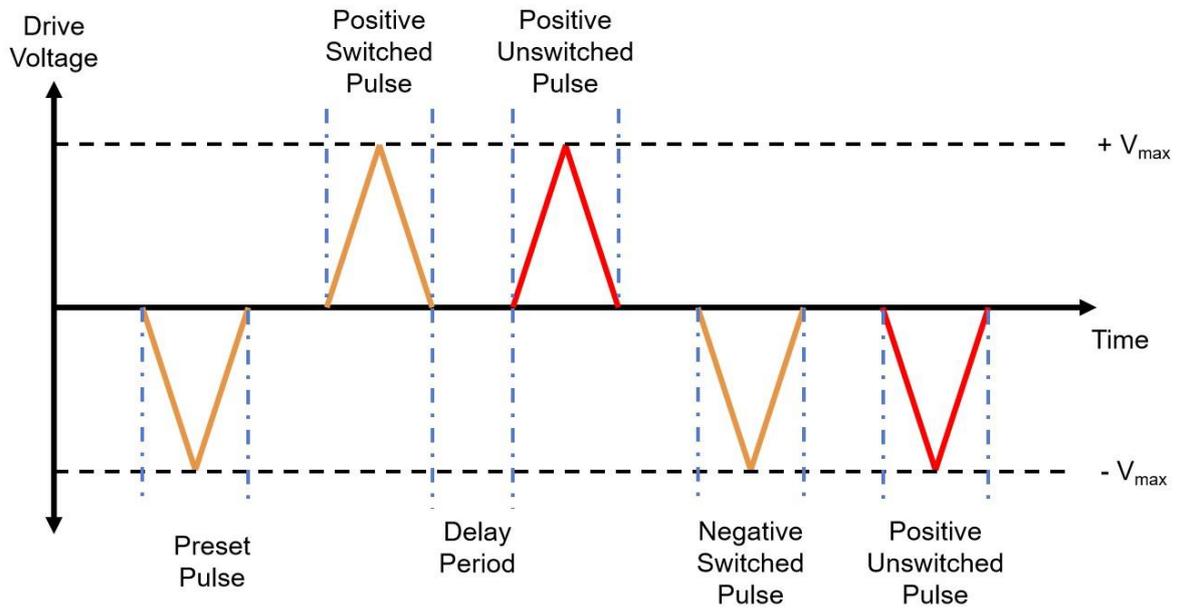

**Figure S3. Positive-up-negative-down (PUND) measurement.** In this experiment, the width and magnitude of the pulses were set to 0.5 ms and 3 MV/cm and the delay period was 1 s.

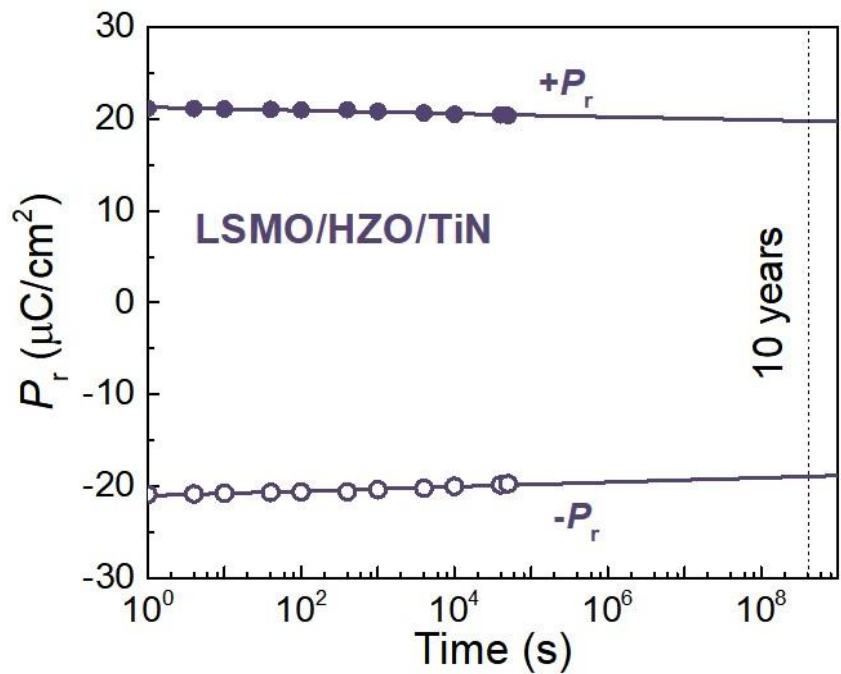

**Figure S4. Retention performance of LSMO/HZO/TiN measured at room temperature.**

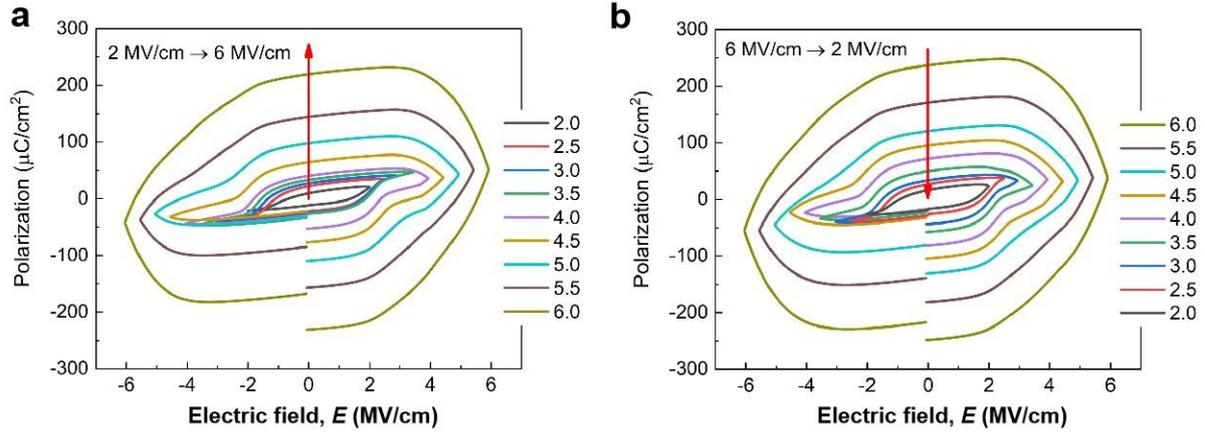

**Figure S5. Polarization of the LSMO/HZO/TiN heterostructure under a series of driving electric fields at 1 kHz. a** *P-E* loops with applied electric field sequential increments from 2 to 6 MV/cm with a step of 0.5 MV/cm. **b** *P-E* loops with an applied electric field decreasing from 6 to 2 MV/cm with a step of 0.5 MV/cm.

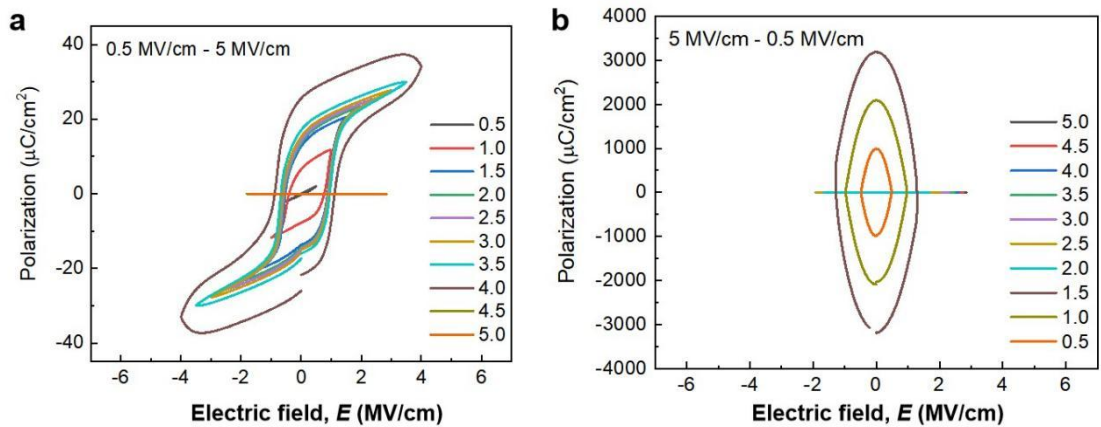

**Figure S6.** *P-E* **loops of the TiN/HZO/TiN heterostructure under a series of driving fields at 1 kHz. a** *P-E* loops with applied electric field sequential increment from 0.5 to 5 MV/cm. **b** *P-E* loops with an applied electric field decreasing from 5 to 0.5 MV/cm. The straight lines in the *P-E* loops show that the leakage current is large, and the integrated polarization exceeds the measuring range of the instrument, indicating that the ferroelectric HZO layer is damaged and breaks down.

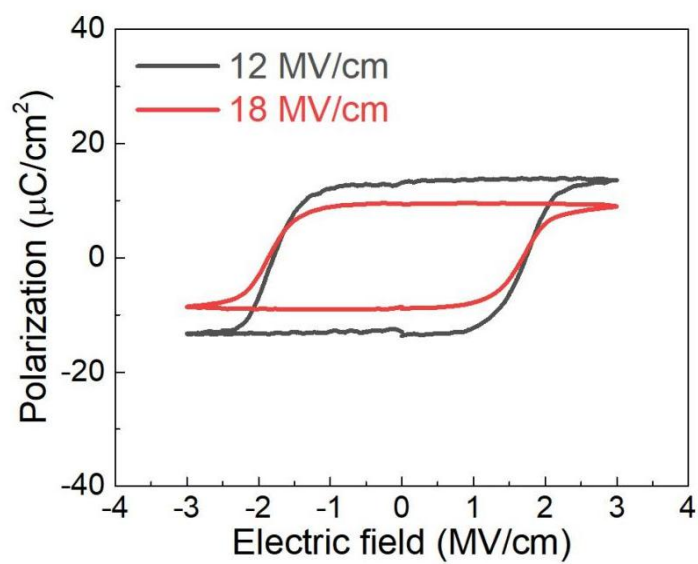

**Figure S7. Polarization of LSMO/HZO/TiN after high electric fields of 12 and 18 MV/cm using the standard PUND method.**

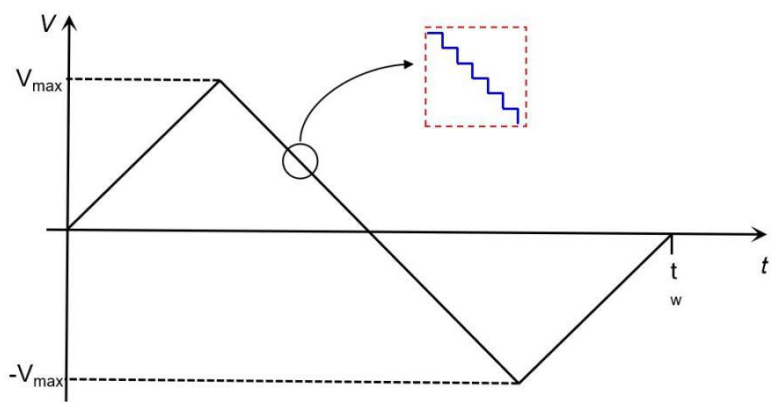

**Figure S8. The standard bipolar triangular waveform for polarization - electric field *(P-E)* and current density - electric field *(J-E)* loops.** $V_{max}$ and $t_w$ can be defined to measure different *P-V* loops at the desired voltage and frequency.

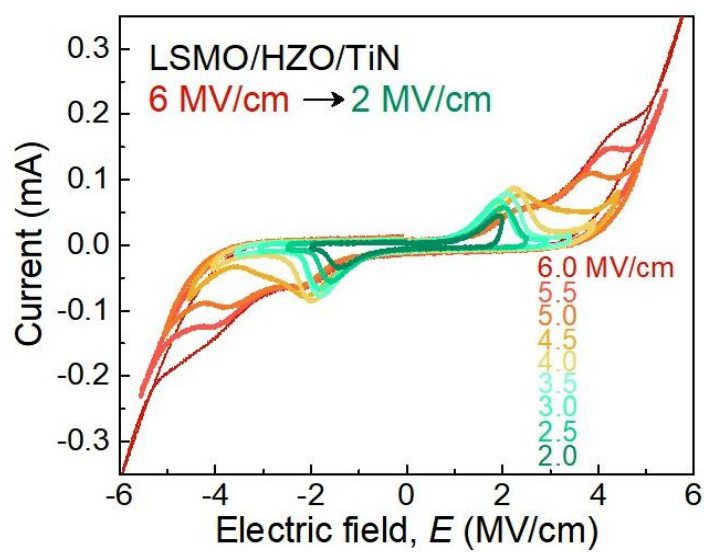

**Figure S9.** *I-E* characteristics of LSMO/HZO/TiN under electric fields ranging from 6 to 2 MV/cm using the standard bipolar triangular waveform method.

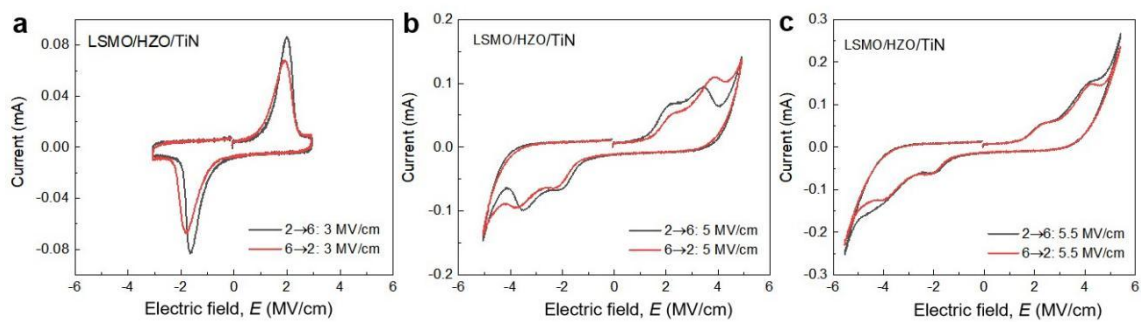

**Figure S10.** *I-E* loops of the LSMO/HZO/TiN heterostructure under the constant driven electric field pulses of (a) 3, (b) 5, and (c) 5.5 MV/cm. The black line and red line in the *I-E* loops represent before (black) and after (red) experiencing an electric field pulse of 6 MV/cm.

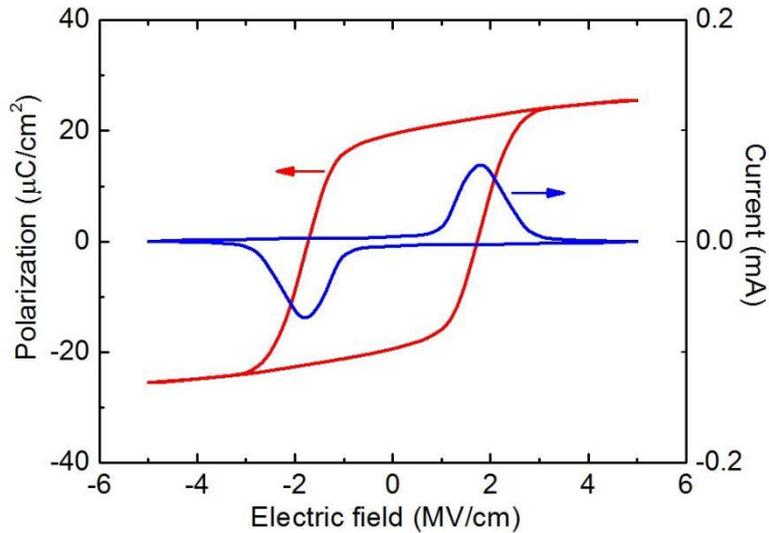

**Figure S11. Simulated *P-E* hysteresis loop and *I-E* curve of LSMO/HZO/TiN with a 15 nm HZO film based on a two-dimensional single-domain phase-field model.** The *P-E* hysteresis loop as a result of the transient currents integrated during a triangular field sweep. Two clear polarization switching current peaks at approximately ±1.9 MV/cm are observed, in agreement with the coercive field of the LSMO/HZO/TiN heterostructure.

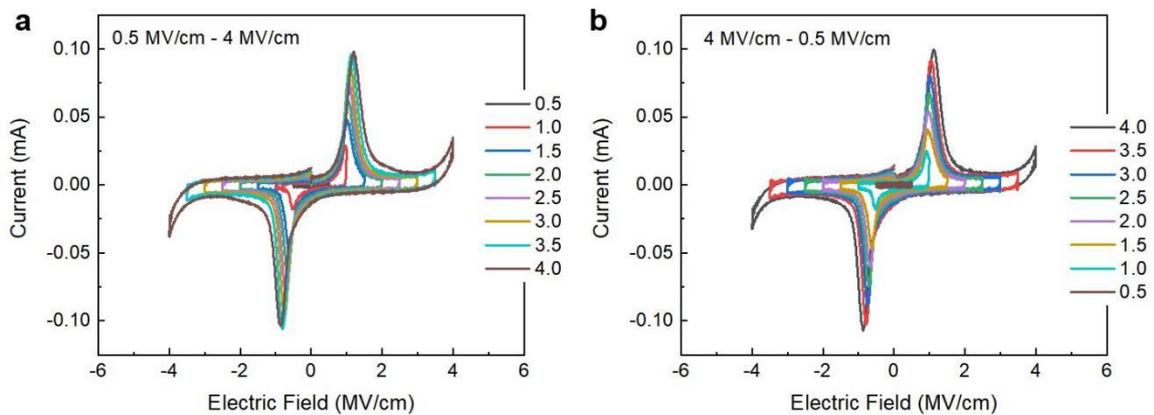

**Figure S12. *I-E* loops of the TiN/HZO/TiN heterostructure under a series of driving fields at 1 kHz. a** *I-E* loops with electric field increments from 0.5 MV/cm to 4 MV/cm. As the driven field increases, the current peak gradually rises, corresponding to the switching of ferroelectric domains. **b** *I-E* loops with electric field decrements from 4 MV/cm to 0.5 MV/cm. When the ferroelectric field decreases, the switching current consequently decreases. Note that *I-E* loops of *E* > 4 MV/cm are absent because the HZO film is totally damaged under such a high electric field.

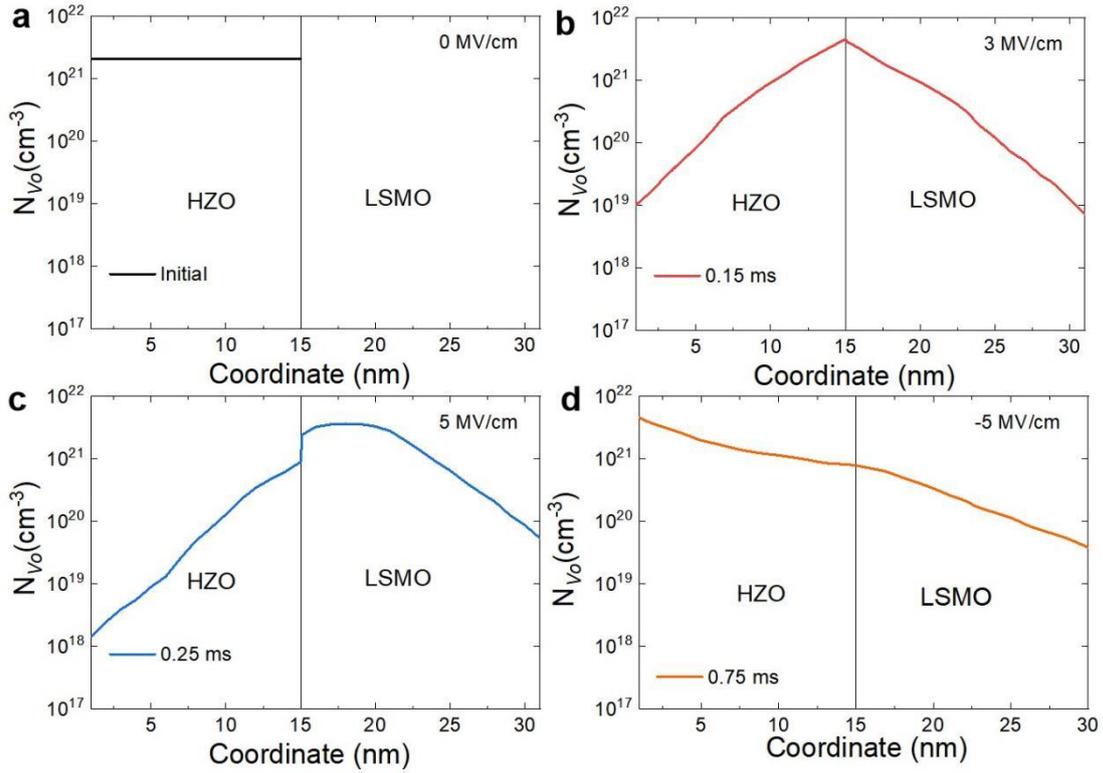

**Figure S13. Simulated oxygen vacancy ($V_O$) concentration profiles in the HZO film and LSMO electrode during the application of a triangular voltage pulse.** Different profiles of the $V_O$ distributions are created by representative triangular voltage pulses, namely (**a**) The initial state; (**b**) 3 MV/cm at 0.15 ms; (**c**) 5 MV/cm at 0.25 ms; and (**d**) −5 MV/cm at 0.75 ms. Because the oxygen ion is negatively charged and has the opposite charge against the oxygen vacancy, the migration direction and distribution of the oxygen ion is the opposite to $V_O$.

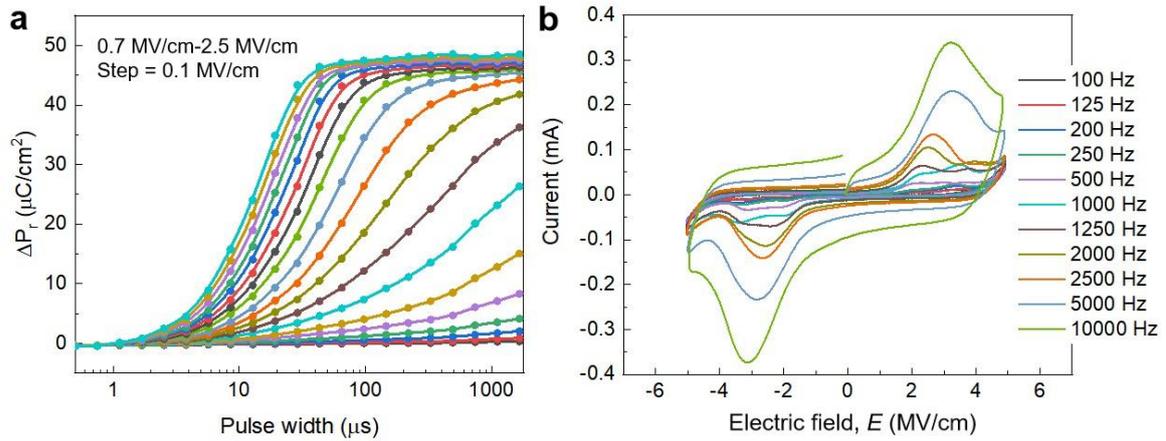

**Figure S14. Dynamic switching property of the LSMO/HZO/TiN heterostructure. a** Dynamic switching speed stimulated by square pulses. A series of switching pulses with alternating pulse widths from 500 ns to 1.5 ms, and changing amplitudes from 0.7 to 2.5 MV/cm with a step of 0.1 MV/cm, were applied on the HZO capacitor. **b** Current-electric field (*I-E*) loops under different field frequencies from 100 Hz to 10 kHz with an amplitude of 5 MV/cm. The current peak shifts to higher electric fields as the field frequency increases (100 Hz to 10 kHz).

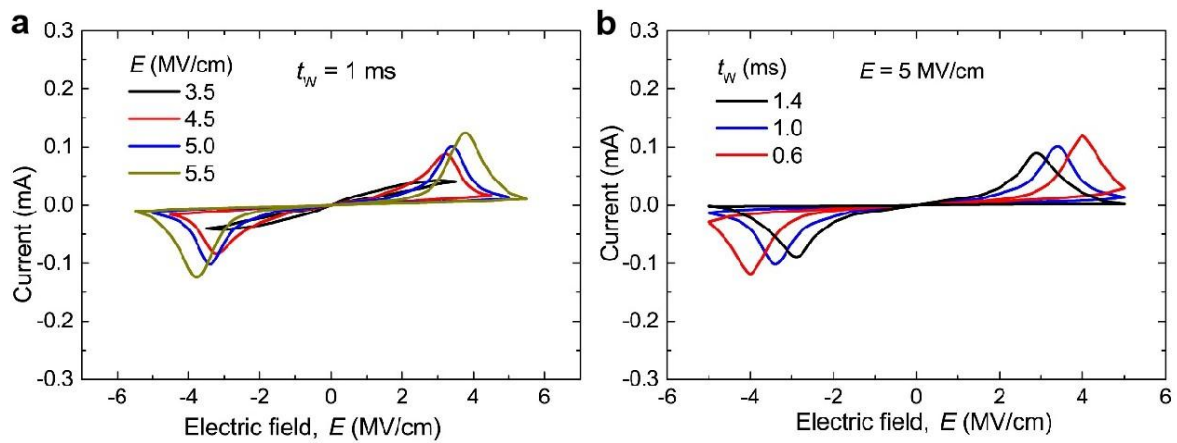

**Figure S15. Simulated current response for electric fields with different magnitudes (a) and sweep rates (b).** A larger electric field (*E*) and faster sweep rate ($t_W$) induce more peaks shifted to the right, which is in agreement with the experimental *I-E* curves.

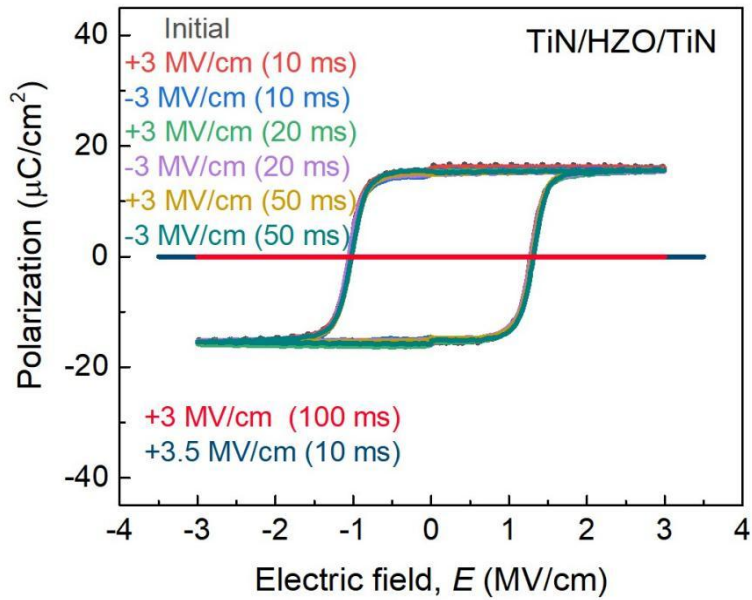

**Figure S16. Polarizations of the TiN/HZO/TiN heterostructure after applying electrical bias fields of ±3 and +3.5 MV/cm under various dc pulse periods.** Under a bias field of ±3 MV/cm with time periods of 10, 20 and 50 ms, the ferroelectric performance of TiN/HZO/TiN remains unchanged. Under 3 MV/cm with time periods approaching 100 ms and 3.5 MV/cm with 10 ms, the TiN/HZO/TiN heterostructures are totally damaged and break down.

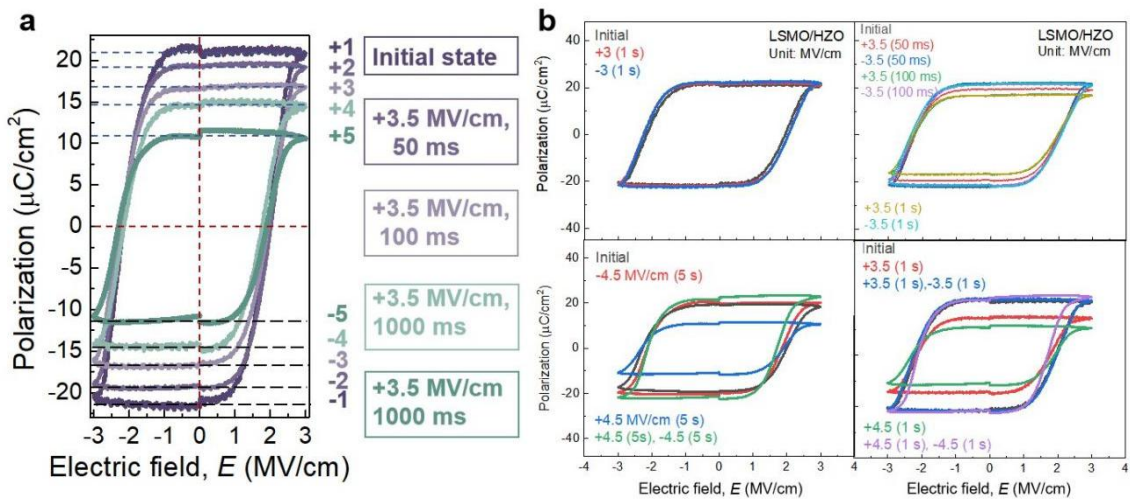

**Figure S17. Multi-level polarization states of the LSMO/HZO/TiN heterostructure after applying various electrical bias fields of different DC pulse periods. a** Positive and negative polarization states are controlled by the electric field. **b** Reproducibility and reversibility of the multiple polarization states in HZO. Under a bias field of ±3 MV/cm with a time period of up to 1 second, compared with the TiN/HZO/TiN heterostructure, the

polarization of LSMO/HZO/TiN remains stable and shows robust ferroelectric performance. Under the bias field of ±3.5 MV/cm with various time periods, LSMO/HZO/TiN heterostructure presents multi-polarization states with reversibility and controllability.

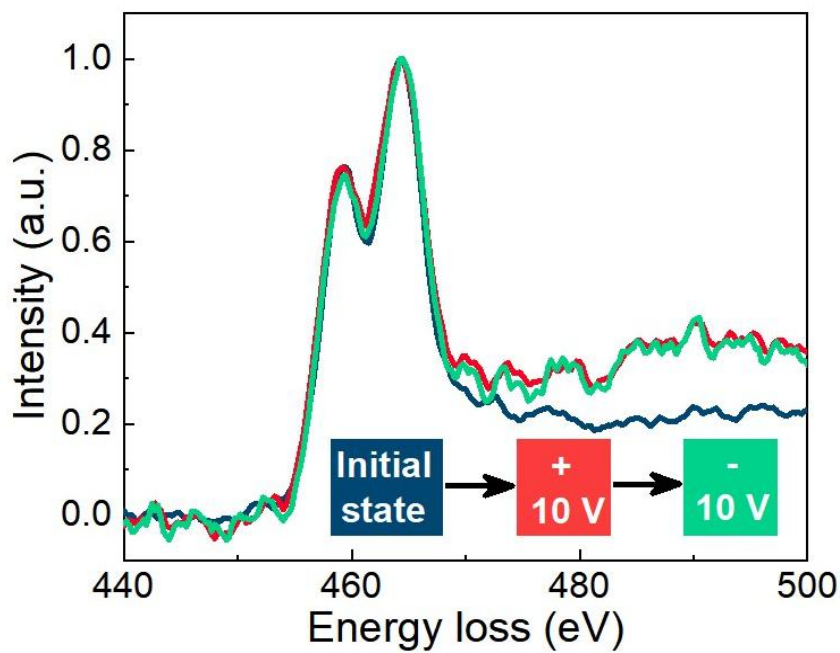

**Figure S18. STEM-EELS of the titanium $L_{2,3}$ edges for the initial state, after positive voltage, and after negative voltage.** STEM-EELS spectra indicate that the chemical environment of TiN was almost unchanged.

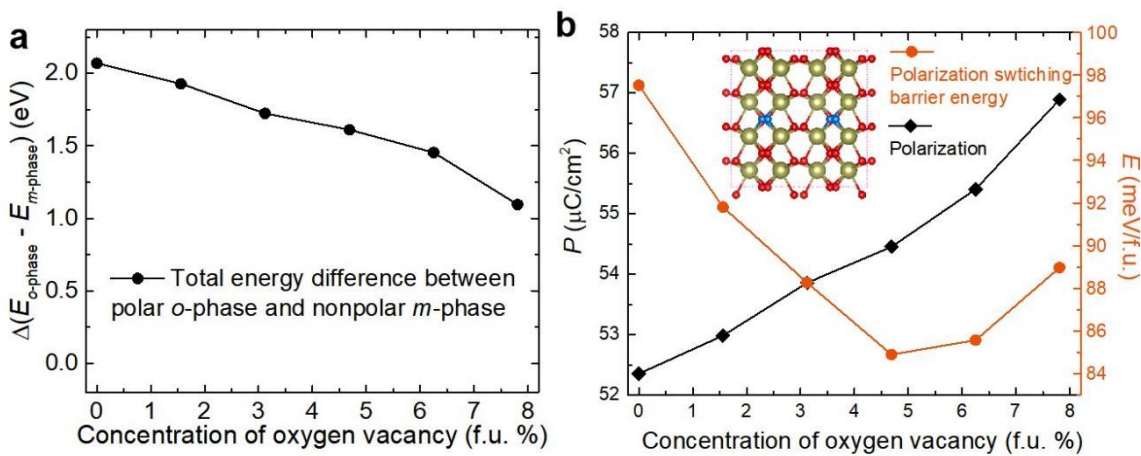

**Figure S19. The effects of oxygen vacancies on the phase stability, spontaneous polarization ($P$) and polarization switching barrier energy ($E$) of HfO$_2$. a** The total energy difference between the $o$-phase and $m$-phase as a function of the vacancy concentrations. **b**

The spontaneous polarization (P) and polarization switching barrier energy (E) of the polar o-phase as a function of the vacancy concentrations.

**Table S1. The parameters used in the phase-field model** (*13*).

| Symbol | value |
| --- | --- |
| $\alpha$ | -3.68 ×10$^8$ V·m/C |
| $\beta$ | -5.21 ×10$^8$ V·m$^5$/C |
| $\gamma$ | -4.67 ×10$^{10}$ V·m$^9$/C |
| $g$ | 1×10$^{10}$ V·m$^3$/C |
| $L$ | 2.3 S/m |